%% file: main.tex
\documentclass[twocolumn]{aastex63}
\usepackage{amsmath}
\usepackage{rotating}

\shorttitle{EXPRES Detection of Fe~{\sc i} on MASCARA-5~b}
\shortauthors{Sikora et al.}

\usepackage{lineno}

\graphicspath{{./}{figures/}}

\defcitealias{petz2025}{P25}

\begin{document}

\title{Confirmation of Fe~{\sc i} on MASCARA-5~b's Dayside Observed With EXPRES}

\correspondingauthor{James T. Sikora}
\email{james.t.sikora@gmail.com}

\author[0000-0002-3522-5846]{James T. Sikora}
\affiliation{Lowell Observatory, 1400 W Mars Hill Road, Flagstaff, AZ 86001, USA}

\author[0000-0003-4450-0368]{Joe Llama}
\affiliation{Lowell Observatory, 1400 W Mars Hill Road, Flagstaff, AZ 86001, USA}

\author[0000-0002-9288-3482]{Rachael M. Roettenbacher}
\affiliation{Department of Astronomy, University of Michigan, Ann Arbor, MI 48109, USA}

\author[0009-0003-8440-5813]{Elisabeth M. Brann}
\affiliation{Department of Astronomy and Planetary Science, Northern Arizona University, Flagstaff, AZ 86011, USA}
\affiliation{Department of Physics, Bryn Mawr College, Bryn Mawr, PA 19010, USA}

\author[0000-0002-0875-8401]{Jean-Michel D\'esert}
\affiliation{Leibniz-Institut für Astrophysik Potsdam (AIP), An der Sternwarte 16, 14482 Potsdam, Germany}
\affiliation{DESY, Platanenallee 6, D-15738 Zeuthen, German}
\affiliation{Anton Pannekoek Institute for Astronomy, University of Amsterdam, 1098 XH Amsterdam, The Netherlands}

\author[0000-0001-7047-8681]{Alex S. Polanski}
\altaffiliation{Percival Lowell Postdoctoral Fellow}
\affil{Lowell Observatory, 1400 W Mars Hill Road, Flagstaff, AZ, 86001, USA}
\affil{Department of Physics and Astronomy, University of Kansas, Lawrence, KS 66045, USA}

\author[0000-0002-7670-670X]{Malena Rice}
\affiliation{Department of Astronomy, Yale University, 219 Prospect St., New Haven, CT 06511, USA}

\author[0000-0002-3852-3590]{Lily Zhao}
\altaffiliation{NASA Sagan Fellow}
\affiliation{Department of Astronomy \& Astrophysics, University of Chicago, 5640 S Ellis Ave, Chicago, IL 60637, USA}

\begin{abstract}
MASCARA-5~b/TOI-1431~b is an ultra-hot Jupiter \citep[$P_{\rm orb}=2.650237\pm0.000003\,{\rm d}$, $T_{\rm eq}=2370\pm70\,{\rm K}$, $M_{\rm p}=3.12\pm0.18\,M_{\rm Jup}$, $R_{\rm p}=1.49\pm0.05\,R_{\rm Jup}$;][]{addison2021} orbiting a bright Am star ($V=8.0\,{\rm mag}$). Recent time-series observations obtained with PEPSI@LBT during the planet's post-eclipse phases have revealed Fe~{\sc i} emission lines indicative of a thermally inverted atmosphere. These observations demonstrate that MASCARA-5~b is well-suited to atmospheric characterization via emission spectroscopy, thereby motivating further follow-up observations covering additional orbital phases to constrain the planet's atmospheric chemistry, thermal structure, and dynamics. Here we present pre-eclipse time-series observations obtained with the high-resolution optical spectrograph EXPRES@LDT. Our analysis confirms the previous detection of gas-phase Fe~{\sc i} on MASCARA-5~b's dayside (with a $5.5\sigma$ significance obtained from two nights of observations) and the fact that the thermal profile is inverted with lower and upper temperatures $\sim2000\,{\rm K}$ and $\sim4500\,{\rm K}$, respectively. A search for Fe~{\sc ii} and Cr~{\sc i} did not yield any plausible detections. We also find that the pre-eclipse signal exhibits a non-negligible blueshift of $-3.2\pm1.4\,{\rm km/s}$ potentially caused by winds.
\end{abstract}


\section{Introduction}\label{sect:intro}

In recent years, high-resolution spectroscopy ($R\gtrsim15,000$) has become a well-established means of studying hot Jupiter phase curves from the ground \citep{birkby2018,vansluijs2023}. In addition to being much less expensive than space-based instruments, the higher resolutions that ground-based instruments are capable of allow individual absorption/emission lines from the planet's atmosphere to be resolved. This is important not only for identifying the various gaseous species that make up the atmosphere's chemical composition \citep[e.g.,][]{brogi2017,hawker2018}, but also because it allows for dynamical information to be extracted, revealing evidence of winds, magnetic drag, and the planet's rotation \citep[e.g.,][]{snellen2014,brogi2016,beltz2022}.

Ultra-hot Jupiters (UHJs) are particularly well-suited to ground-based high-resolution spectroscopy both at near infrared (IR) and optical wavelengths \citep{hoeijmakers2018a,kasper2021,brogi2023} due to their exceptionally high equilibrium temperatures ($T_{\rm eq}\gtrsim2000\,{\rm K}$). Nearly all UHJs with constrained pressure-temperature profiles exhibit temperature inversions in which the upper atmosphere is heated to extremely high temperatures \citep{petz2025}. This, in turn, can lead to high abundances of H$^{\rm -}$ \citep{arcangeli2018,jacobs2022} and ionized metals \citep[e.g., Fe~{\sc ii}, Ca~{\sc ii}, Cr~{\sc ii}, Mg~{\sc ii};][]{hoeijmakers2019,yan2019,stangret2020} where the latter may impact atmospheric flow patterns through complex magnetic effects \citep{perna2010,rogers2014,beltz2022,christie2025,blocker2026}. With the development of high-resolution cross-correlation (HRCC) techniques as applied to exoplanet atmospheric characterization \citep[e.g.,][]{snellen2010,birkby2013,brogi2019,pino2020,kesseli2022}, high-resolution ground-based observations are now providing critical tests of our understanding of how hot and ultra-hot Jupiter atmospheric chemistry, thermal structure, and dynamics are linked to one another and their role in terms of planet formation processes \citep{line2021,pelletier2023,smith2024a}.

MASCARA-5~b/TOI-1431~b/HD~201033~b is an UHJ orbiting a bright non-magnetic metallic-line A-type star ($V=8.0\,{\rm mag}$, spectral type Am) \citep{renson1991}. It has a $2.650237\pm0.000003\,{\rm d}$ orbital period, a high $T_{\rm eq}$ of $2370\pm70\,{\rm K}$, and a mass and radius of $3.12\pm0.18\,M_{\rm Jup}$ and $1.49\pm0.05\,R_{\rm Jup}$ inferred by \citet{addison2021} based on ground-based photometry, high-resolution spectroscopy/radial velocity measurements, and TESS photometry. The planet's secondary eclipse and full phase curve were detected in the TESS photometry implying high dayside and nightside temperatures of $T_{\rm day}=3004\pm64\,{\rm K}$ and $T_{\rm night}=2583\pm63\,{\rm K}$ \citep{addison2021}. No stellar variability (e.g., rotational modulation) in the TESS light curve was reported by these authors who note that the star's projected rotational velocity ($v\sin{i}=6.0\pm0.2\,{\rm km/s}$) and radius ($R_{\rm \star}=1.92\pm0.07\,R_\odot$) implies a rotation period $\sim10-16\,{\rm d}$. The planet's Rossiter-McLaughlin effect was previously detected through high-resolution optical spectra obtained with the EXtreme PREcision Spectrograph (EXPRES) implying a mis-aligned orbit ($\lambda=-155_{-10}^{+20}\,{\rm degrees}$) \citep{stangret2021}; however, the same in-transit observations did not reveal any atmospheric absorption lines, which may be explained by the planet's relatively high surface gravity and low atmospheric scale height. Recently, \citet{petz2025} (hereafter \citetalias{petz2025}) reported a $5.68\sigma$ detection of Fe~{\sc i} emission lines (along with a tentative detection of Cr~{\sc i}) on the planet's dayside using a single night of PEPSI@LBT observations obtained during the post-eclipse phases ($\phi=0.507-0.594$). Additional pre-eclipse observations ($\phi=0.447-0.481$) did not reportedly yield any robust detections of the atmosphere, which may be due to the lower S/N ($\sim150$) and shorter sequence length (27 exposures) compared to the post-eclipse observations (${\rm S/N}\sim400$ and 68 exposures).

In this study, we apply HRCC methods to time-series EXPRES observations obtained during MASCARA-5~b's pre-eclipse phase in order to detect and characterize the planet's previously detected emission spectrum. These new observations are highly complementary to the previous PEPSI observations considering the extended phase coverage and the similar bandpasses/resolutions of the two instruments. Our analysis confirms the presence of gaseous Fe~{\sc i} on the planet's dayside and confirms the atmosphere's inverted thermal structure. In Sect. \ref{sect:obs}, we describe the EXPRES observations along with the data reduction process. The cleaning steps applied to the data in order to isolate the planetary signal and the framework used for the subsequent atmospheric retrieval are described in Sect. \ref{sect:analysis}. Our results are presented in Sect. \ref{sect:results} followed by a discussion and our conclusions in Sect. \ref{sect:discussion}.

\begin{figure}
	\centering
    \includegraphics[width=1\columnwidth]{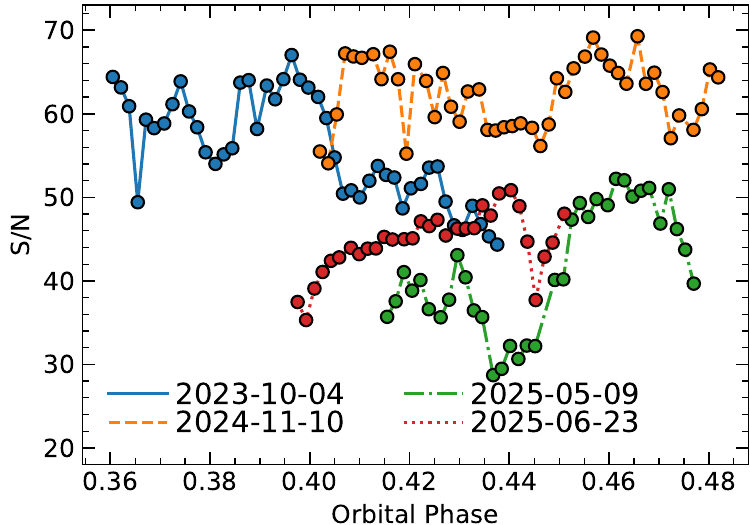}
	\caption{S/N near $5520\,$\r{A} and orbital phase coverage for the four pre-eclipse EXPRES observations. The eclipse ingress starts at a phase of $0.48$.}
	\label{fig:obs_snr}
\end{figure}

\section{EXPRES Spectra}\label{sect:obs}

Time-series high-resolution spectra of MASCARA-5 were obtained using the EXPRES instrument installed on the $4.3\,{\rm m}$ Lowell Discovery Telescope \citep{levine2012}. EXPRES is an \'echelle spectrograph with a median resolving power of $137,500$, a wavelength range of $3800-8220\,$\r{A} \citep{jurgenson2016,blackman2020}, and is equipped with both a Laser Frequency Comb (LFC) and ThAr lamp that are used for wavelength calibration. The data were reduced using the EXPRES pipeline \citep{petersburg2020}, which is based on flat-relative optimal extraction \citep{zechmeister2014}. Barycentric corrections are applied pixel-by-pixel thanks to a chromatic exposure meter \citep{blackman2017}. EXPRES spectra are continuum normalized by iteratively fitting a B-spline with asymmetric sigma-clipping, where outliers with values less than the model are discounted. Tellurics are modeled using SELENITE \citep{leet2019}, which makes use of an empirical model of changing telluric lines built from a database of EXPRES B-star spectra.

We observed MASCARA-5 over four nights coinciding with the planet's pre-eclipse phase ($\phi\approx0.36-0.48$) on Oct. 4 2023 UT, Nov. 10 2024 UT, May 9 2025 UT, and June 23 2025 UT. A total of 46 consecutive $350\,{\rm sec}$ exposures were taken on each of the first two nights over an $\approx5\,{\rm hr}$ duration while 35 and 31 exposures were obtained during the last two nights over $3.9\,{\rm hr}$ and $3.4\,{\rm hr}$ durations, respectively. The airmasses during nights one and two increased from $1.1$ near the start of the observations to $1.4$ and $1.8$ at the end of each sequence, respectively. For nights three and four, the airmasses decreased from $2.0$ to $1.1$ and from $1.2$ to $1.1$ between the start and end of each sequence, respectively. Clouds were not reported during any of the four sequences; the lower S/N for nights three and four may be attributed to poor/variable seeing. A ThAr and LFC spectrum was taken at the start and end of each sequence and also after every fifth exposure (approximately every $30\,{\rm min}$) to enable generating a precise wavelength solution. We used the ThAr wavelength solution in our analysis considering the wider wavelength coverage ($3797-8224\,$\r{A} compared to $4982-7265\,$\r{A}). The S/N near $5520\,$\r{A} and orbital phase coverage associated with each night of observations is shown in Fig. \ref{fig:obs_snr}.

\begin{figure}
	\centering
    \includegraphics[width=1\columnwidth]{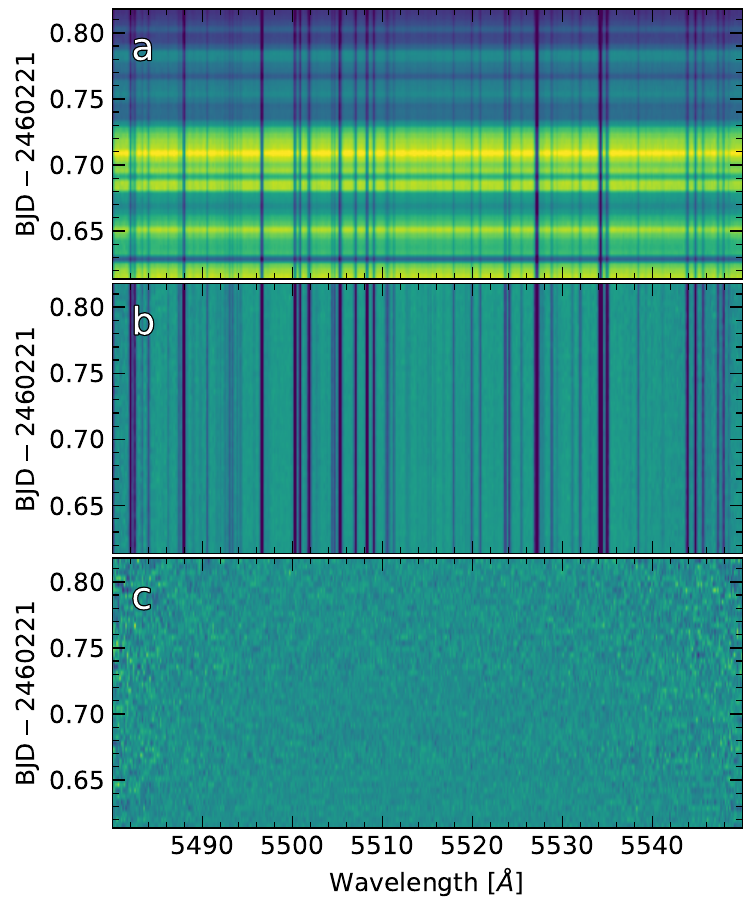}
	\caption{An example of the cleaning/detrending steps applied to the EXPRES observations after telluric correction. Panel `a' shows the continuum-normalized flux from the EXPRES reduction pipeline, panel `b' shows the median-normalized flux, and panel `c' shows the final cleaned spectra after applying the \texttt{SYSREM} algorithm and a median box filter.}
	\label{fig:cleaning}
\end{figure}

\section{Analysis}\label{sect:analysis}

\subsection{Cleaning \& Detrending}\label{sect:cleaning}

The first step of the analysis involved applying a cleaning pipeline to the reduced and continuum-normalized spectra to effectively remove the stellar absorption lines, telluric lines (i.e., the residual tellurics that were not fully removed during the data reduction noted in the previous section), and any significant instrumental/systematic variability. Each night of observations was grouped into a 3D data cube with dimensions of $N_{\rm order}\times N_{\rm exp}\times N_{\rm pixel}$, corresponding to the number of spectral orders (86), the number of exposures taken during a single night (31-46), and the number of pixels/wavelength channels in each spectral order (7920), respectively. Three masking criteria were then applied. Similar to \citet{hoeijmakers2020}, we masked the outer $5\%$ of pixels on either edge of each order (784/7920 pixels in total), where the photon count is naturally low due to the blaze response typical of \'echelle spectrographs. We also masked six orders that show strong contamination from telluric lines (orders 72, 76, 77, 80, 81, and 86). Finally, we masked any columns/lightcurves with one or more pixels that have $>5\sigma$ outlier flux values (calculated relative to the median and standard deviation of each column). Next, we shifted each spectrum into the stellar rest frame using the radial velocities (RVs) calculated from each spectrum, which allows for the stellar absorption lines to be removed more effectively. The RVs were derived using a Gaussian fit to the cross-correlation function (CCF) between the EXPRES spectra and a line list tuned for A0 stars \citep{petersburg2020}. The best fit mean and covariance of the Gaussian fit are taken to be the RV and corresponding RV error. For the first two higher-S/N data sets, the median RV error is $2.5\,{\rm m/s}$ while the last two lower-S/N data sets have a median error of $4\,{\rm m/s}$.

Order-by-order detrending in wavelength and time was then carried out starting with an initial normalization of each row/spectrum in the $N_{\rm exp}\times N_{\rm pixel}$ subarray by that row's median flux. This serves to remove the bulk of the flux variations caused by changes in the seeing/conditions that occur between exposures. We then applied the \texttt{SYSREM} iterative detrending algorithm \citep{tamuz2005,mazeh2007} as implemented in \texttt{petitRADTRANS} (v3) \citep{molliere2019,blain2024b}. We found that three iterations of the algorithm removed most of the stellar contribution to the Fe~{\sc i} cross-correlation function (described in Sect. \ref{sect:cc}). Order-by-order sigma clipping was then carried out on the detrended spectra using 10 iterations in which pixels in each 2D $N_{\rm exp}\times N_{\rm pixel}$ subarray having $>4\sigma$ outlier flux values (calculated using the full 2D array) were identified and masked. Lastly, a median box filter with a width of 151 pixels was calculated from each spectrum using \texttt{scipy}'s \texttt{median\_filter} function and subsequently subtracted in order to remove any residual large-scale variations. An example of the cleaning steps carried out for the first night (2023-10-04) is shown in Fig. \ref{fig:cleaning} for order $\#50$ centered on $5520\,$\r{A}.

Several alternative detrending methods were tested and compared with the results obtained using the three-iteration \texttt{SYSREM}-based detrending approach described above. Using a higher number of \texttt{SYSREM} iterations was not found to significantly change either the amplitude of stellar residuals or the significance of the detected Fe~{\sc i} signal presented in Sect. \ref{sect:detect}: we found that the detection significance obtained by combining nights one and two increases from $<4\sigma$ when using one iteration to a maximum of $5.5\sigma$ when using three iterations and subsequently varying between $5.0\sigma$ and $5.4\sigma$ when using up to ten iterations. Additionally, we tried applying the \texttt{polyfit} detrending algorithm included in \texttt{petitRADTRANS} and described by \citet{blain2024b}, we explored the impact of using more aggressive or alternative masking criteria \citep[e.g.,][]{cabot2019}, and we tried using additional smoothing/normalizing procedures applied before or after \texttt{SYSREM} or principal component analysis is carried out \citep[e.g.,][]{pelletier2023}. No alternative methods yielded any clear improvement in terms of the magnitude of the stellar residuals found in the CCF or in terms of the detection significance. We note that our search for planetary signals is not expected to be significantly impacted by the presence of stellar residuals given the large differences in velocities between the planet and host-star during nearly all exposures (discussed further in Sect. \ref{sect:results}).

\begin{figure}
	\centering
    \includegraphics[width=1\columnwidth]{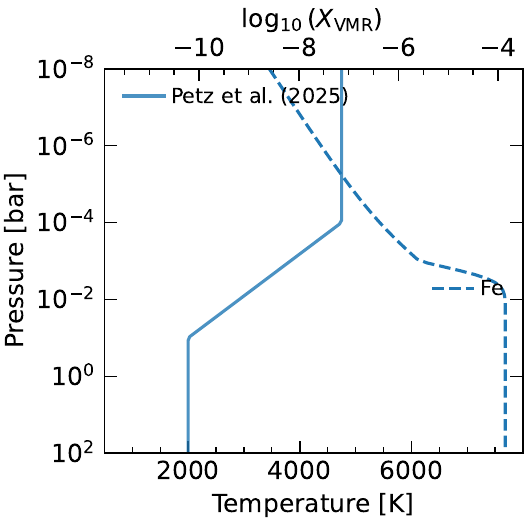}
	\caption{MASCARA-5~b's atmospheric temperature (bottom axis) and Fe~{\sc i} VMR (top axis) adopted for the model emission spectrum. The solid blue line is the PT profile, which is based on that used by \citetalias{petz2025} to detect Fe~{\sc i} with PEPSI@LBT. The dashed blue line shows the Fe~{\sc i} VMRs calculated using \texttt{FastChem} assuming chemical equilibrium and a stellar metallicity of ${\rm [M/H]}=0.09$.}
	\label{fig:PT}
\end{figure}

\begin{figure*}
	\centering
    \includegraphics[width=1\textwidth]{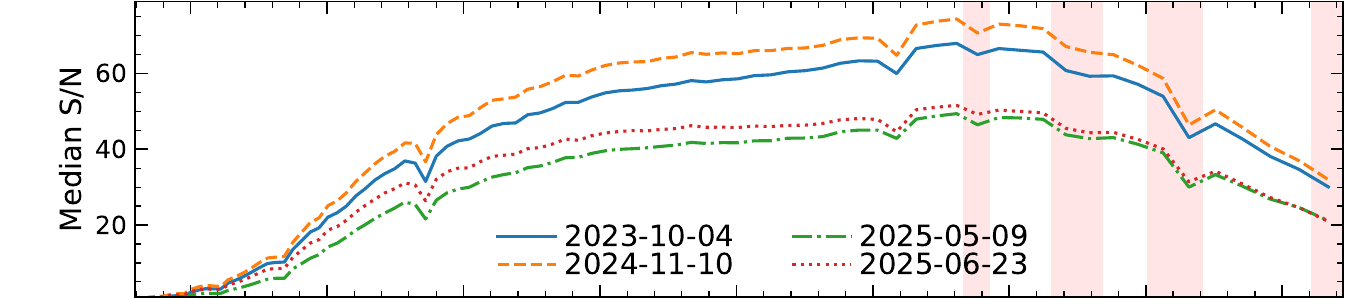}
    \includegraphics[width=1\textwidth]{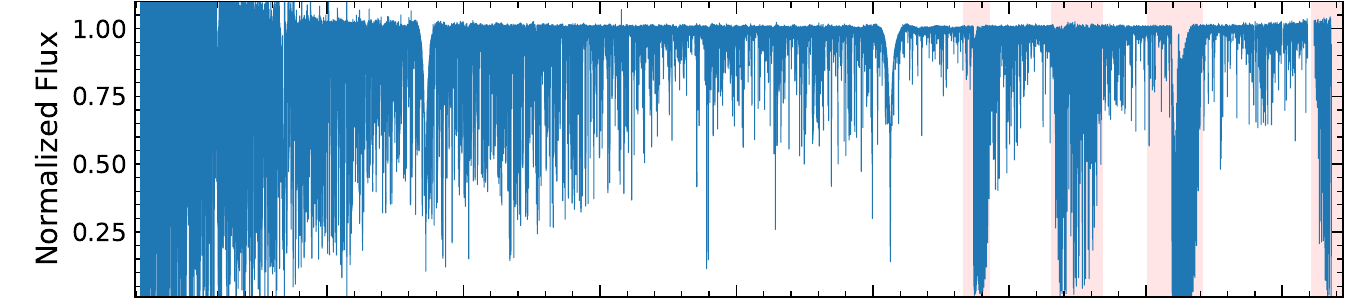}
    \includegraphics[width=1\textwidth]{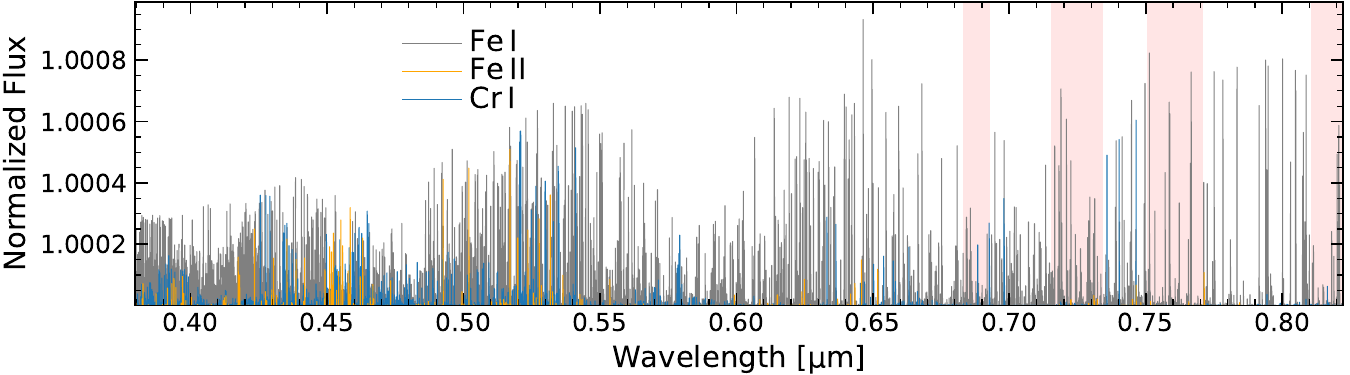}
	\caption{\emph{Top:} Median observed S/N for each of the 86 EXPRES spectral orders. \emph{Middle:} Median observed spectrum normalized by the stellar continuum and including tellurics. \emph{Bottom:} Continuum-normalized model templates used in this study. We show templates calculated for Fe~{\sc i} (detected in our study with a $5.1\sigma$ significance) along with Fe~{\sc ii} and Cr~{\sc i} (not detected). Red shaded regions indicate orders that were masked due to heavy telluric contamination.}
	\label{fig:template}
\end{figure*}

\subsection{Cross-Correlation}\label{sect:cc}

We searched for evidence of the planet's emission spectrum by cross-correlating the cleaned EXPRES data with model emission spectra calculated using the radiative transfer code \texttt{petitRADTRANS} (v3) \citep{molliere2019,blain2024b}. The code can produce high-resolution ($R=10^6$) model spectra for a specified pressure-temperature (PT) profile and atmospheric chemical composition. To save computational time, we reduced the line-by-line opacity sampling resolution by a factor of two yielding model emission spectra ($F_{\rm p}[\lambda]$) with a resolution of $R=500,000$. The model atmosphere was constructed using 100 pressure layers logarithmically spaced from $10^2-10^{-8}\,{\rm bar}$. For the PT profile, we used the 3-layer profile proposed by \citet{brogi2014}, which consists of upper and lower isothermal layers that bound an intermediate layer having a temperature gradient (or that may be isothermal if the upper and lower isothermal layers have equal temperatures). It is defined by an upper boundary (near the top of the atmosphere) occurring at temperatures and pressures of $T_1$ and $P_1$, respectively, and a lower boundary at $T_2$ and $P_2$ such that the intermediate layer's temperature gradient is
\begin{equation}
    T_{\rm slope}=\frac{T_1-T_2}{\log_{10}P_{\rm 1}-\log_{10}P_{\rm 2}}
\end{equation}
\citep{yan2020}. Chemical abundances at each pressure layer were calculated assuming chemical equilibrium using \texttt{FastChem} (v3.1.2) \citep{stock2018} for a given metallicity and C/O value.

In order to account for the loss of the relative continuum flux, we continuum-normalized the model spectra following the approach outlined by \citet{yan2020}: $F_{\rm p}(\lambda)$ is divided by the stellar blackbody spectrum \citep[$F_{\rm s}(\lambda)$, calculated assuming $T_{\rm eff}=7690\,{\rm K}$;][]{addison2021} and then normalized by the planet's continuum flux given by the blackbody flux of the lower pressure layer \citep[i.e., where the temperature is defined by $T_2$; see Fig. B.2 of][]{yan2020}. Instrumental broadening was applied to the normalized model spectra using a Gaussian kernel (using the \texttt{instrBroadGaussFast} function from the \texttt{pyasl}/PyAstronomy Python package\footnote{\url{https://pyastronomy.readthedocs.io/en/latest/index.html}}) assuming an instrumental resolution of $140,000$. Rotational broadening was then added using the \texttt{pyasl.fastRotBroad} function with no limb darkening and a rotational velocity of $2.9\,{\rm km/s}$, which assumes that the planet is tidally locked. Fig. \ref{fig:template} shows continuum-normalized and broadened model emission spectra calculated by including either gas-phase Fe~{\sc i}, Fe~{\sc ii}, or Cr~{\sc i}. The models are generated using an approximation of the PT profile adopted by \citetalias{petz2025} such that $T_1=4750\,{\rm K}$, $P_1=10^{-3}\,{\rm bar}$, $T_2=2000\,{\rm K}$, and $P_2=10^{-1}\,{\rm bar}$ (Fig. \ref{fig:PT}) and assuming a stellar metallicity of ${\rm [M/H]}=0.09$. Note that this is an approximation due to the different PT profile models used where \citetalias{petz2025} use a \citet{guillot2010}-based profile (see their Fig. 2) that cannot be exactly replicated by the model adopted in our study. 

\begin{figure*}
	\centering
    \includegraphics[width=1\columnwidth]{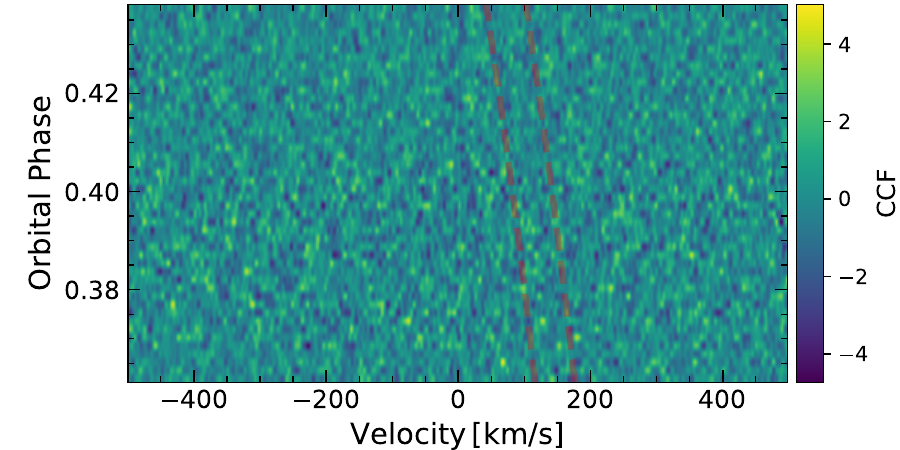}
    \includegraphics[width=1\columnwidth]{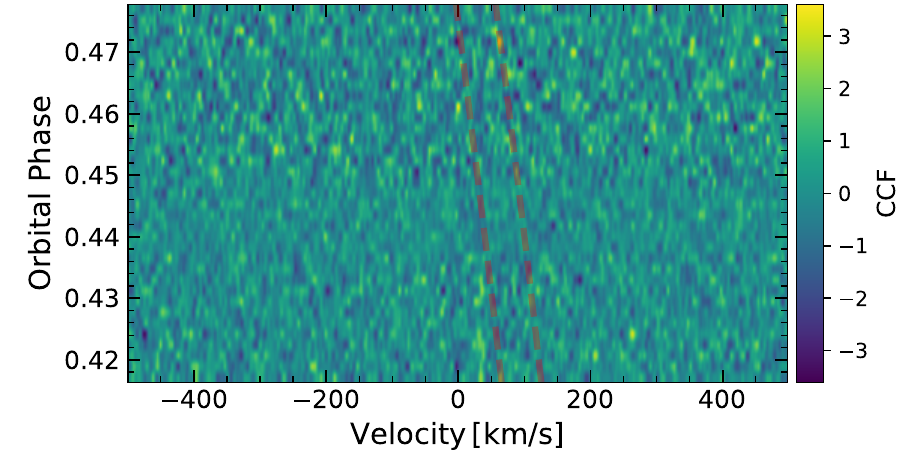}
    \includegraphics[width=1\columnwidth]{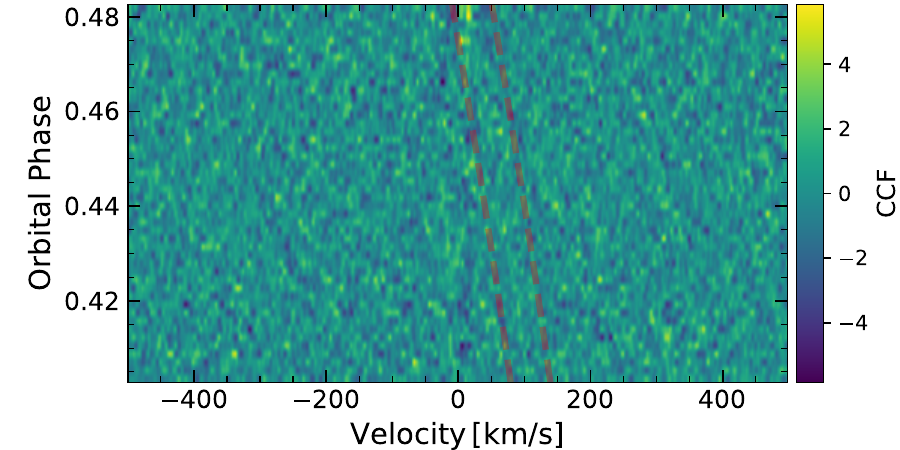}
    \includegraphics[width=1\columnwidth]{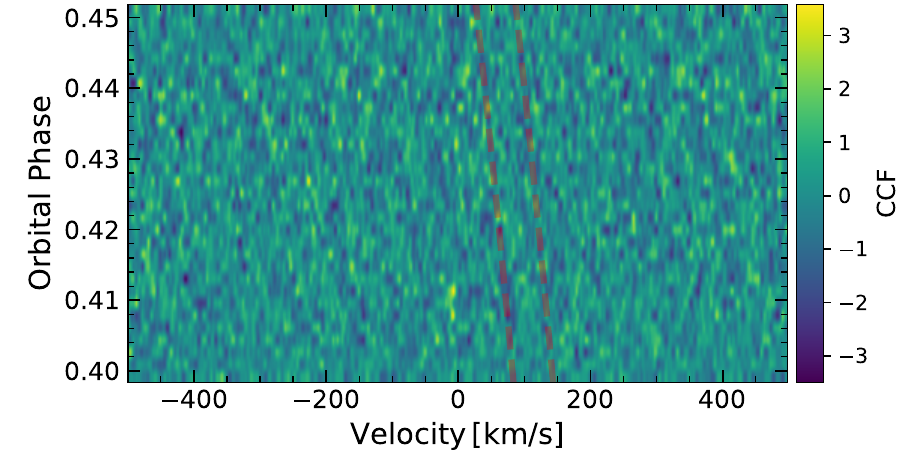}
	\caption{Stellar rest-frame ${\rm CCF}(v,t)$ values calculated using the Fe~{\sc i} template (Fig. \ref{fig:template}) for the four nights of observations: 2023-10-04 (top left), 2024-11-10 (bottom left), 2025-05-09 (top right), and 2025-06-23 (bottom right). The red dashed lines bracket the expected planetary radial velocity assuming $K_{\rm p}=187.6\,{\rm km/s}$ and $\Delta v=0\,{\rm km/s}$.}
	\label{fig:CCF}
\end{figure*}

The \texttt{SYSREM} detrending pipeline distorts any planet spectrum that may be present in the observations. We accounted for this distortion using the method described by \citet{gibson2022} in which we compute the $\textbf{U}$ and $\Lambda$ matrices (i.e., the matrices consisting of the \texttt{SYSREM} basis vectors and the measurement uncertainties, respectively) and applying the model transformation given by their Eqn. 7.

The search for atmospheric signals was carried out by first applying the \texttt{SYSREM} matrix transformation to the templates shown in Fig. \ref{fig:template} and then cross-correlating with the cleaned EXPRES data using the expression adopted by \citet{gibson2020}:
\begin{equation}\label{eqn:ccf}
    {\rm CCF}(v,t)=\sum\frac{f(t)g(v)}{\sigma^2(t)}
\end{equation}
where $f(t)$ is the observed spectrum at time $t$, $g(v)$ is the model template, Doppler-shifted by velocity $v$ to match the planet's radial velocity, and $\sigma(t)$ are the measurement errors. The planet velocity is calculated assuming a circular orbit:
\begin{equation}\label{eqn:v_planet}
    v_{\rm p}(t)=K_{\rm p}\sin(2\pi\phi)+\Delta v
\end{equation}
where $K_{\rm p}$ is the planet's radial velocity semi-amplitude, $\phi$ is the orbital phase calculated using the parameters published by \citet{addison2021} (listed in Table \ref{tbl:param} below), and $\Delta v$ is the radial velocity offset, which may be non-zero as previously found in other HRCC hot Jupiter studies \citep[e.g.,][]{snellen2010,line2021}.

For each spectrum (i.e., for every night, exposure, and order), ${\rm CCF}(v,t)$ was calculated using a model template that was Doppler-shifted over a velocity grid spanning $-400\,{\rm km/s}$ to $+400\,{\rm km/s}$ in $1\,{\rm km/s}$ intervals. We then integrated ${\rm CCF}(v,t)$ along planet radial velocities defined over a grid of $K_{\rm p}$ and $\Delta v$ values ($-250\,{\rm km/s}<K_{\rm p}<250\,{\rm km/s}$ and $-150\,{\rm km/s}<\Delta v<150\,{\rm km/s}$). Strong peaks in the integrated CCF (${\rm CCF}[K_{\rm p},\Delta v]$, also known as a $K_{\rm p}-\Delta v$ or $K_{\rm p}-V_{\rm sys}$ map) found near the expected $K_{\rm p}\approx187\,{\rm km/s}$ and $\Delta v\approx0\,{\rm km/s}$ are then interpreted as a potential detection of the planet's atmosphere. Such peaks can be be converted into a detection significance or S/N by normalizing the ${\rm CCF}(K_{\rm p},\Delta v)$ map by the standard deviation with some studies opting to first mask the region containing the detected planet signal \cite[e.g.,][]{birkby2013,brogi2014}. We opted to use the full ${\rm CCF}(K_{\rm p},\Delta v)$ map when normalizing by the standard deviation.

\subsection{Markov-Chain Monte Carlo Sampling}\label{sect:mcmc}

The detected Fe~{\sc i} signal (presented in Sect. \ref{sect:results}) was characterized using an MCMC approach in conjunction with model atmosphere templates calculated with \texttt{petitRADTRANS}. MCMC sampling was carried out using \texttt{emcee} \citep{foreman-mackey2019} in which six walkers were generated, 35,000 steps in length, with the first 5,000 steps being discarded as burn-in. Convergence was tested by (1) comparing the ratio of the total chain length to the autocorrelation length ($N_{\rm steps}/\tau_{\rm auto}$), and (2) using the Gelman-Rubin convergence statistic \citep[$\hat{R}$,][]{gelman1992}. We concluded that sufficient convergence had been achieved for chains that are $30,000$ steps in length based on the fact that, for each free parameter, $N_{\rm steps}/\tau_{\rm auto}\geq50$ and $\hat{R}<1.001$. For the log-likelihood function, we used that proposed by \citet{gibson2020} and given by
\begin{equation}
    \log L=-\frac{N}{2}\log\left(\frac{1}{N}\sum_{i=1}^{N}\frac{(f_i-a g_i)^2}{\sigma_i^2}\right),
\end{equation}
where $f$, $\sigma$, and $g$ are the measured flux values, measurement uncertainties, and the model flux values (as defined in Eqn. \ref{eqn:ccf}), respectively, $N$ is the total number of data points, and $a$ scales the continuum-normalized model template (e.g., Fig. \ref{fig:template}) in order to adjust the line strengths that largely depend on the temperature gradient and the chemical abundances. The forward model used in our analysis consists of three free parameters: $K_{\rm p}$ and $\Delta v$, which parameterize the velocity of the planetary signal, and $a$ (parameterized as $\log_{10}a$). All other parameters (e.g., $T_1$, $T_2$, $\log_{10}P_1$, and $\log_{10}P_2$, metallicity) were fixed. This was done due to the fact that only Fe~{\sc i} was detected in our analysis, which, without the presence of additional detected species, can lead to poor parameter constraints \citep{yan2020}. We confirmed this by carrying out initial atmospheric retrievals, which showed strong degeneracies between the PT profile parameters and the metallicity/Fe abundance. Moreover, \citetalias{petz2025} have previously detected Fe~{\sc i} via the planet's dayside emission spectrum thereby providing a suitable initial guess for MASCARA-5~b's atmospheric thermal structure.

\section{Results}\label{sect:results}

\subsection{Neutral Fe Detection}\label{sect:detect}

In Fig. \ref{fig:CCF}, we show the ${\rm CCF}(v,t)$ values calculated using the Fe~{\sc i} model template for each of the four nights of observations. A residual stellar signal is visible by eye near $v\approx0\,{\rm km/s}$ in many of the exposures, particularly for the two lower-S/N nights. Both the Fe~{\sc ii} and Cr~{\sc i} ${\rm CCF}(v,t)$ values show similar, relatively weak stellar residuals. In all cases, no individual exposure shows clear evidence of a planetary signal.

\begin{figure*}
	\centering
   \includegraphics[width=0.95\columnwidth]{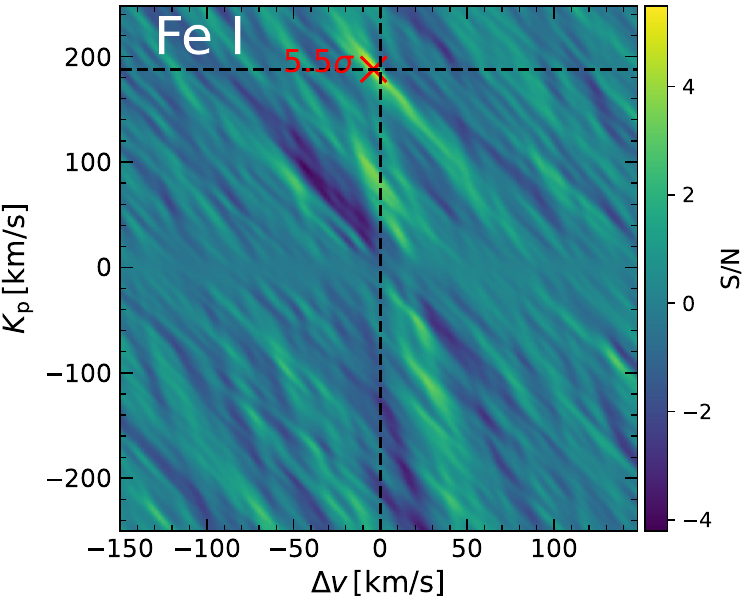}
   \includegraphics[width=0.95\columnwidth]{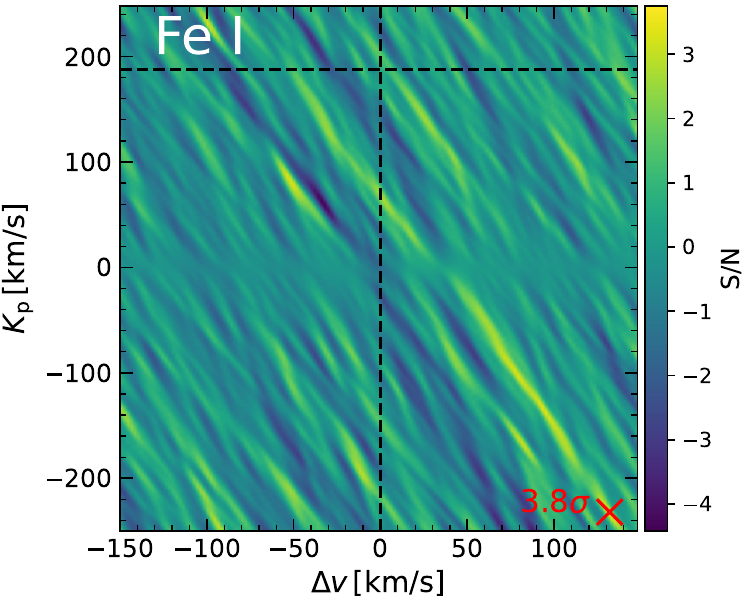}
   \includegraphics[width=0.95\columnwidth]{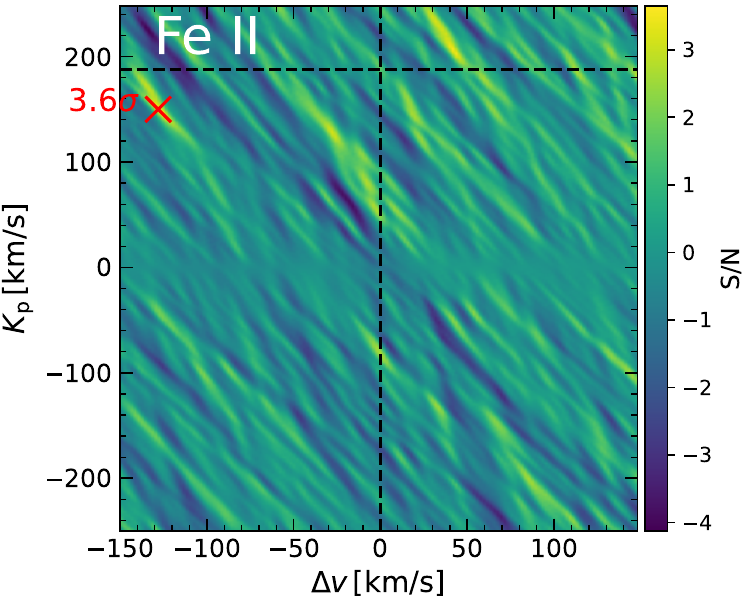}
   \includegraphics[width=0.95\columnwidth]{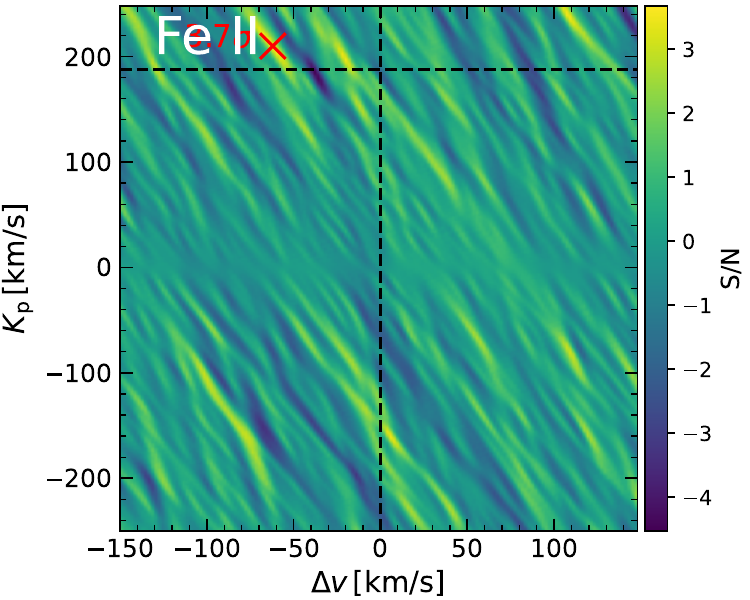}
   \includegraphics[width=0.95\columnwidth]{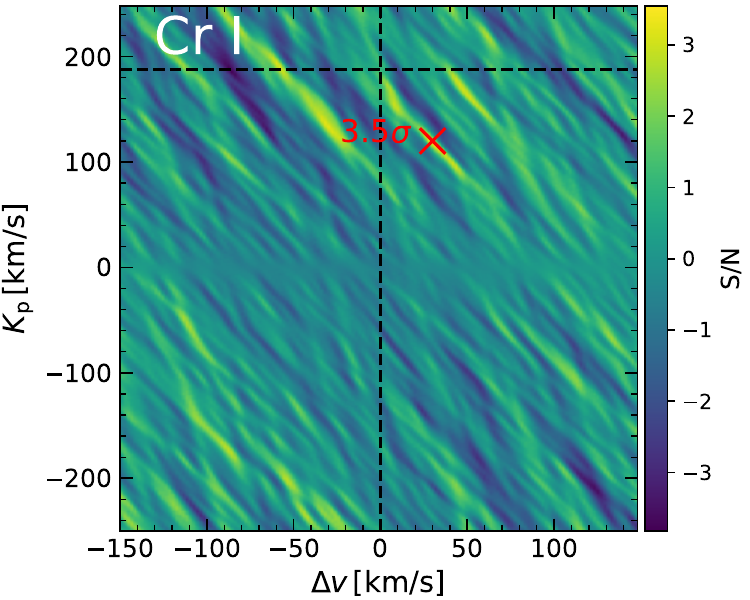}
   \includegraphics[width=0.95\columnwidth]{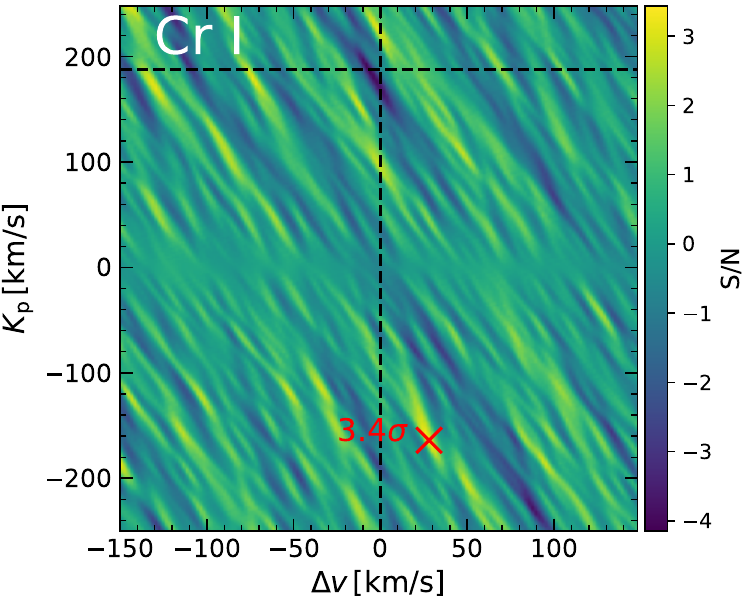}
	\caption{Normalized, stellar rest-frame ${\rm CCF}(K_{\rm p},\Delta v)$ maps calculated using the model templates generated for Fe~{\sc i} (top row), Fe~{\sc ii} (middle row), or Cr~{\sc i} (bottom row). The left and right columns correspond to the combination of the first two nights (2023-10-04 and 2024-11-10) and the last two nights (2025-05-09 and 2025-06-23). The dashed black lines indicate the expected $K_{\rm p}=187.6\,{\rm km/s}$ and $\Delta v=0\,{\rm km/s}$. The red `x' symbols mark the maximum S/N. Fe~{\sc i} is the only detected species in this study ($5.3\sigma$, top left).}
	\label{fig:KpVsys}
\end{figure*}

In the case of the integrated ${\rm CCF}(K_{\rm p},\Delta v)$ maps calculated for Fe~{\sc i}, we find that each of the first two, higher-S/N nights (2023-10-04 and 2024-11-10) show maximum-amplitude peaks that occur near the expected $K_{\rm p}\approx187\,{\rm km/s}$ and $\Delta v\approx0\,{\rm km/s}$ and have $>3.5\sigma$ significance. The last two, lower-S/N nights (2025-05-09 and 2025-06-23) exhibit maximum-amplitude peaks at $K_{\rm p}$ and $\Delta v$ that are entirely inconsistent with the expected values. In Fig. \ref{fig:KpVsys}, we show the ${\rm CCF}(K_{\rm p},\Delta v)$ maps calculated when combining nights 1 and 2 (left column) or nights 3 and 4 (right column) when using an Fe~{\sc i} template (first row). Combining nights 1 and 2 reveals a $5.5\sigma$ Fe~{\sc i} detection with $K_{\rm p}\approx187\,{\rm km/s}$ and $\Delta v\approx-3\,{\rm km/s}$ while combining nights 3 and 4 shows no peaks that can plausibly be attributed to a planetary origin with peak S/N values $<4\sigma$. In the case of Fe~{\sc ii} and Cr~{\sc i} (middle and bottom rows of Fig. \ref{fig:KpVsys}), we find no evidence of plausible planetary signals.

\begin{figure}
	\centering
    \includegraphics[width=1\columnwidth]{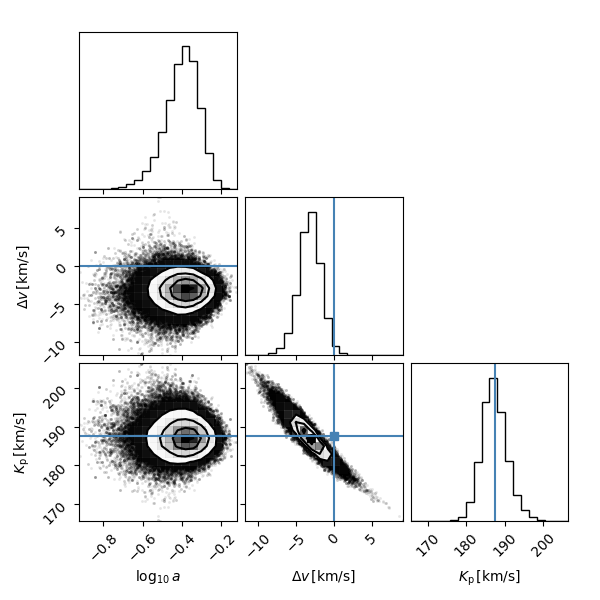}
	\caption{Marginalized posterior distributions obtained from the MCMC sampling analysis using the Fe~{\sc i} forward model. The blue lines indicate the expected values of $K_{\rm p}=187.6\,{\rm km/s}$ and $\Delta v=0\,{\rm km/s}$.}
	\label{fig:corner}
\end{figure}

Stellar contamination can produce statistically significant peaks in the integrated ${\rm CCF}(K_{\rm p},\Delta v)$ maps, particularly for those exposures in which the difference between the velocities of the star and planet are low. In our data, the expected planet velocity assuming $K_{\rm p}=187.6\,{\rm km/s}$ and $\Delta v=0\,{\rm km/s}$ (bracketed by the red dashed lines in Fig. \ref{fig:CCF}) is generally largely separated from the stellar velocity with the minimum, median, and maximum velocity differences across all exposures being $22$, $85$, and $147\,{\rm km/s}$. We tested the extent to which the ${\rm CCF}(K_{\rm p},\Delta v)$ maps are impacted by stellar contamination by first masking all ${\rm CCF}(v,t)$ values at $|v|<20\,{\rm km/s}$. These maps (presented in Fig. \ref{fig:KpVsys_vlim20} of the Appendix) show a higher Fe~{\sc i} detection significance for the combination of nights 1 and 2 ($6.2\sigma$) while all other instances exhibit comparable peaks with amplitudes $\leq4.1\sigma$. We also found that combining all 4 nights resulted in a lower Fe~{\sc i} detection significance whether or not the ${\rm CCF}(v,t)$ values at $|v|<20\,{\rm km/s}$ were masked. Based on this result and on that of the injection-recovery tests described below, we opted to consider only the first two nights of observations that show evidence of Fe~{\sc i} in the subsequent MCMC analysis described in Sect. \ref{sect:mcmc}.

As is evident from Fig. \ref{fig:KpVsys}, the detected Fe~{\sc i} signal shows a slight blueshift $\sim-3\,{\rm km/s}$. The MCMC analysis used to further quantify this offset yielded a $K_{\rm p}=187.1\pm3.4\,{\rm km/s}$ and a $\Delta v=-3.2\pm1.4\,{\rm km/s}$. For the model template scaling parameter, we obtain $\log_{10}{a}=-0.39_{-0.09}^{+0.08}$ suggesting that the observed Fe~{\sc i} line strengths are lower than those associated with the model (e.g., due to a lower temperature gradient or lower Fe abundance). As noted by \citet{gibson2020}, the scaling parameter may also be biased to lower values, which the authors mitigate by using measurement uncertainties calculated from the observed count rates. We tested this by re-running the Fe~{\sc i} injection-recovery test MCMC sampling analysis using uncertainties derived from the method described by \citet{gibson2020}. We find that, relative to the result obtained using the pipeline uncertainties, the marginalized $\log_{10}{a}$ distribution is shifted to slightly higher values ($\log_{10}{a}=-0.37\pm0.06$), however, the impact is relatively small. The derived parameters are listed alongside the published systemic parameters in Table \ref{tbl:param} while the corner plot showing the marginalized posterior distributions is shown in Fig. \ref{fig:corner}.

\input{tables/published_param_table}

\subsection{Welch's $t-$test}

\begin{figure}
	\centering
   \includegraphics[width=0.95\columnwidth]{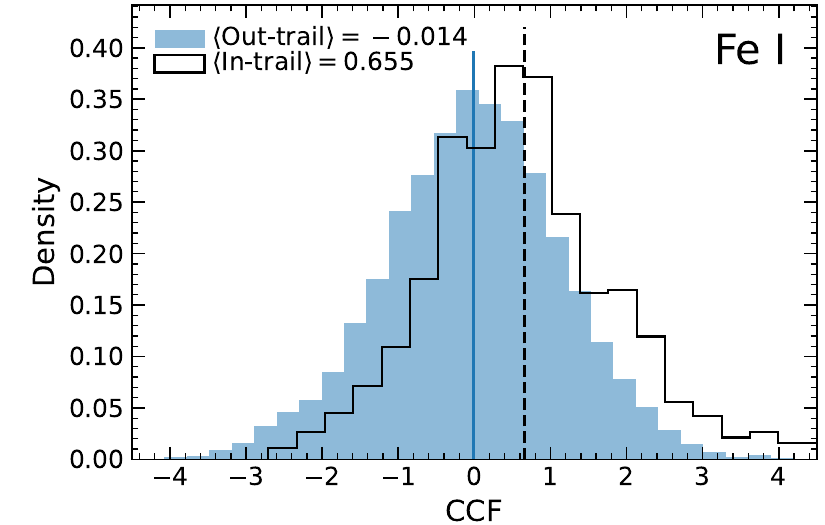}
	\caption{Distributions of ${\rm CCF}(v,t)$ values obtained using the Fe~{\sc i} model template and by combining the data sets from nights 1 and 2. The filled blue and black outlined distributions are the out-of-trail values and the in-trail values, respectively. The Welch's $t-$test rejects the null hypothesis (i.e., that the two samples are drawn from equal-means samples) at a $10\sigma$ significance.}
	\label{fig:welch}
\end{figure}

The Welch's $t-$test, which can be used to test the null hypothesis that two samples are drawn from populations having equal means, is often used to validate a detected cross-correlation signal \citep[e.g.,][]{birkby2013,birkby2017,kanumalla2024}. This involves comparing samples of ${\rm CCF}(v,t)$ values found inside and outside the planet trail (see Fig. \ref{fig:CCF}); assuming that the out-of-trail values are dominated by Gaussian noise, they are expected to be centered on zero while, for a detected signal, the distribution of in-trail values should have a net-positive shift (a mean $>0$). We applied this to the combined nights 1 and 2 data sets for the Fe~{\sc i} ${\rm CCF}(v,t)$ values. The in/out-of-trail distributions were constructed by first shifting each ${\rm CCF}(v,t)$ array into the planetary rest-frame (defined by $K_{\rm p}=187.1\,{\rm km/s}$ and $\Delta v=-3.2\,{\rm km/s}$, based on the detected Fe~{\sc i} signal) and then selecting the in-trail values to be those with $|v|<2.9\,{\rm km/s}$ (i.e., the predicted planet rotational velocity) and selecting the out-of-trail values to be those with $(25<v<125)\,{\rm km/s}$, which do not contain any obvious stellar residuals. The resulting distributions are compared in Fig. \ref{fig:welch} in which we find mean values of $0.655$ and $-0.014$ for the in- and out-of-trail distributions, respectively. Applying the $t-$test results in a rejection of the null hypothesis---that the two samples are drawn from equal-means samples---at a $10\sigma$ significance level.

\subsection{Injection-Recovery Tests}\label{sect:inject}

We also validated the detected Fe~{\sc i} signal (along with the non-detections) by carrying out injection-recovery tests. This involved injecting each of the continuum-normalized model templates (Fe~{\sc i}, Fe~{\sc ii}, and Cr~{\sc i}) shown in Fig. \ref{fig:template} into the EXPRES data itself, starting from the reduced, continuum-normalized spectra with wavelengths defined in the stellar rest frame. For each exposure, the injected model was first Doppler-shifted using the calculated $v_p(t)$ (Eqn. \ref{eqn:v_planet}) with $\Delta v=0\,{\rm km/s}$ and with a negative $K_{\rm p}$ of $-190.4\,{\rm km/s}$. A $\Delta v$ value of $0\,{\rm km/s}$ was chosen since we are simply interested in knowing whether the forward model and MCMC sampling can accurately constrain any velocity offset. Using a negative $K_{\rm p}$ prevents the injected planet spectrum from directly overlapping with the detected spectrum. After injecting the signal, the same cleaning/detrending approach described in Sect. \ref{sect:cleaning} was applied to the data set.

\begin{figure*}
	\centering
   \includegraphics[width=0.95\columnwidth]{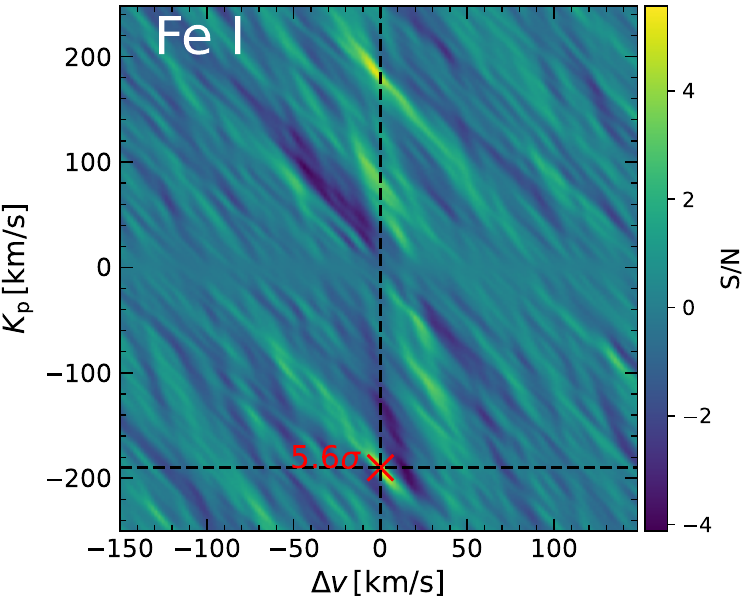}
   \includegraphics[width=0.95\columnwidth]{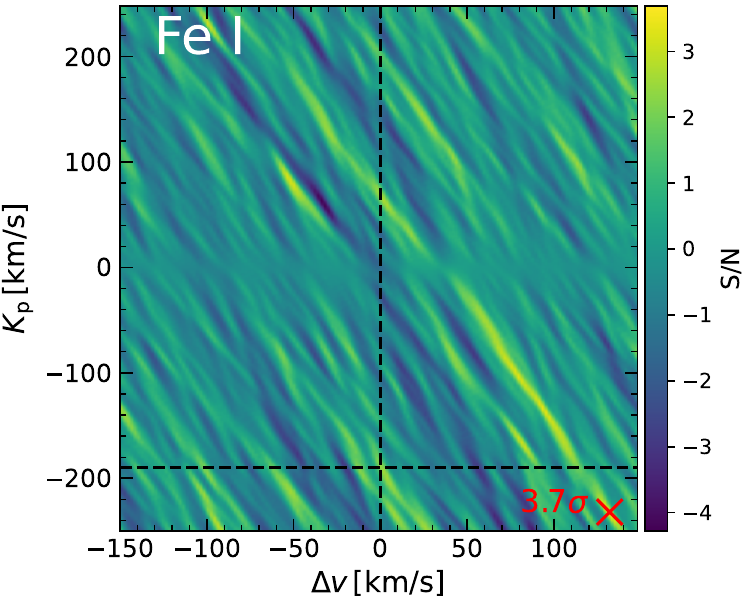}
   \includegraphics[width=0.95\columnwidth]{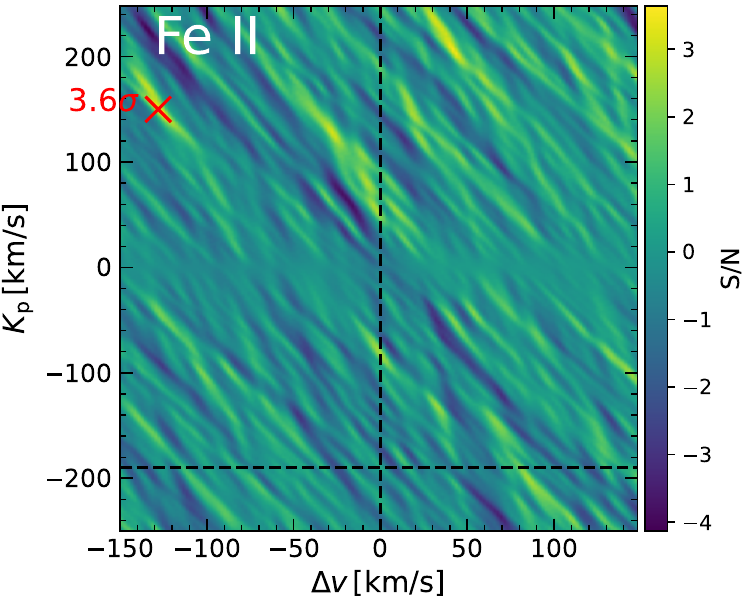}
   \includegraphics[width=0.95\columnwidth]{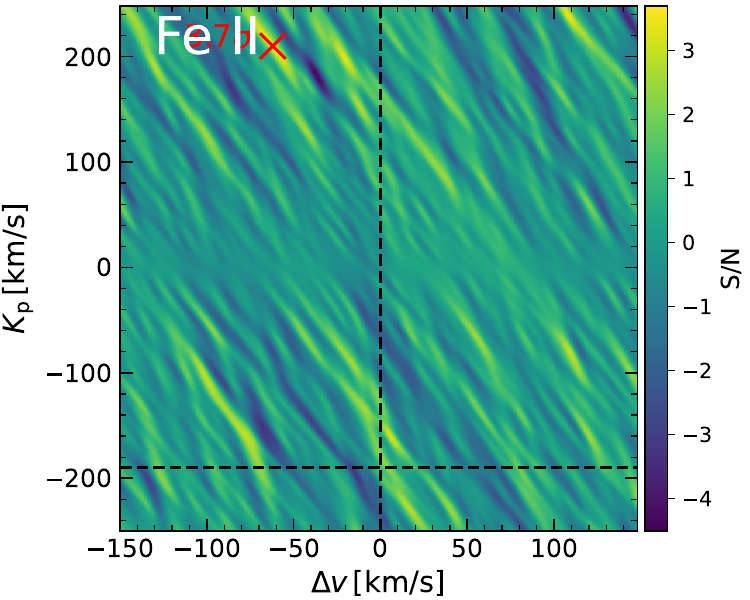}
   \includegraphics[width=0.95\columnwidth]{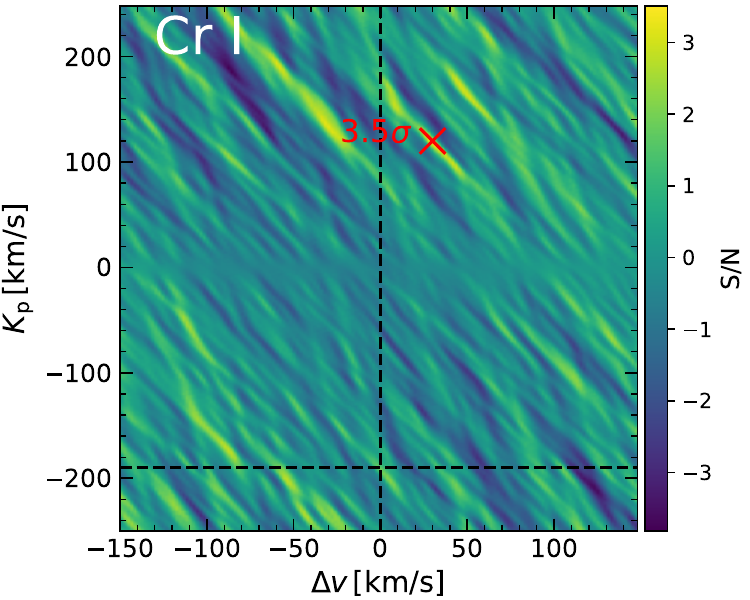}
   \includegraphics[width=0.95\columnwidth]{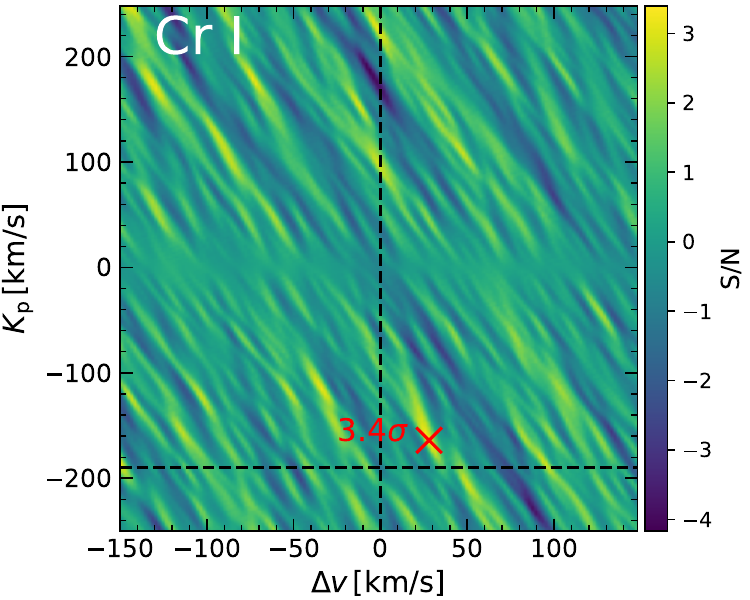}
	\caption{Same as Fig. \ref{fig:KpVsys} but including injected signals with $K_{\rm p}=-190.3\,{\rm km/s}$. Only the Fe~{\sc i} signal injected into the first two nights of observations yielded a clear detection ($5.6\sigma$) with all other cases having S/N$\leq3.7\sigma$}
	\label{fig:KpVsys_inject}
\end{figure*}

The cross-correlation analysis was then carried out on each data set with the injected signals in order to generate ${\rm CCF}(K_{\rm p},\Delta v)$ maps similar to those shown in Fig. \ref{fig:KpVsys}. Combining the first two nights of data containing the injected Fe~{\sc i} templates, we obtain a high ${\rm S/N}=10.6\sigma$, which is significantly higher than the detected signal (${\rm S/N}=5.5\sigma$). However, scaling the injected signal using $\log_{10}a=-0.3$ yields a more comparable S/N of $5.6\sigma$. We therefore scale all three of the injected templates using $\log_{10}a=-0.3$. In Fig. \ref{fig:KpVsys_inject}, we show the ${\rm CCF}(K_{\rm p},\Delta v)$ maps for the injection-recovery tests. We find that, aside from the Fe~{\sc i} case shown on the top left, all of the maps have maximum peaks that are far removed from the injected $K_{\rm p}$ and $\Delta v$ values and have S/N$\leq3.7\sigma$. The injection-recovery tests are therefore consistent with the $5.5\sigma$ Fe~{\sc i} detection being real and non-spurious. The $3.4\leq{\rm S/N}\leq3.8\sigma$ peaks found in the non-detection cases are considered spurious and therefore provide an estimate of the detection threshold.

\begin{figure}
	\centering
    \includegraphics[width=1\columnwidth]{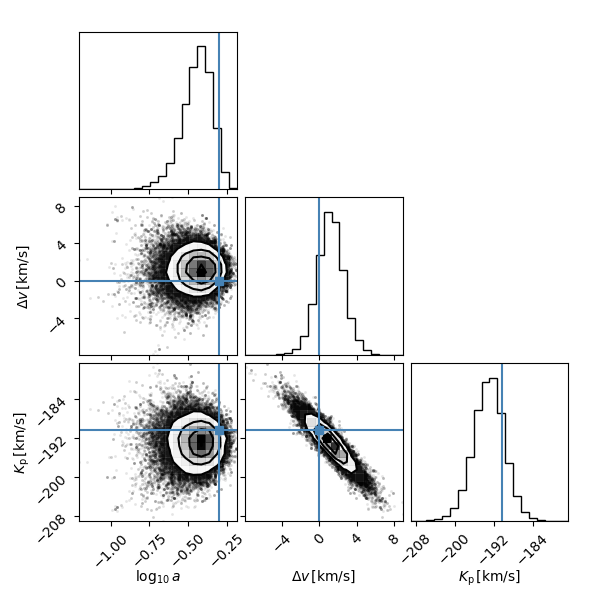}
	\caption{Marginalized posterior distributions obtained from the MCMC sampling analysis injection-recovery testing. The blue lines indicate the values of the injected signal with $K_{\rm p}=-190.3\,{\rm km/s}$, $\Delta v=0\,{\rm km/s}$, and $\log_{10}a=-0.3$.}
	\label{fig:inject_corner}
\end{figure}

Next, we carried out the same MCMC analysis that was used for the detected Fe~{\sc i} signal presented in Sect. \ref{sect:detect}. The resulting marginalized posteriors are shown in Fig. \ref{fig:inject_corner}, which yield $K_{\rm p}=-192.9\pm3.2\,{\rm km/s}$, $\Delta v=1.2\pm1.4\,{\rm km/s}$, and $\log_{10}a=-0.43_{-0.11}^{0.08}$. These values are listed in Table \ref{tbl:inject} alongside the values used for the injected signals. We find $K_{\rm p}$ and $\Delta v$ show negligible $0.8\sigma$ and $0.9\sigma$ discrepancies with respect to the injected values. The scaling parameter constraint has a slightly larger $1.6\sigma$ discrepancy relative to the injected signal.

\input{tables/inject_table}

\section{Discussion and Conclusions}\label{sect:discussion}

Considering MASCARA-5~b's bright stellar host ($V=8.0\,{\rm mag}$), the high planetary equilibrium temperature \citep[$T_{\rm eq}=2370\pm70\,{\rm K}$;][]{addison2021}, it is clearly well-suited to more in-depth atmospheric characterization via emission spectroscopy. Our results confirm previous findings of \citetalias{petz2025}, which detected the presence of gas-phase Fe~{\sc i} on the planet's dayside at a $5.68\sigma$ significance level (based on a single night of post-eclipse LBT observations) along with an inverted PT profile. Our analysis of the pre-eclipse EXPRES observations also finds that MASCARA-5~b's detected Fe~{\sc i} signal exhibits a notable blueshift of $-3.2\pm1.4\,{\rm km/s}$ while the derived $K_{\rm p}$ value ($187.1\pm3.4\,{\rm km/s}$) is in close agreement with the expected value ($187.6\pm3.3\,{\rm km/s}$) that we calculated from the published $M_\star$, $M_{\rm p}$, $P_{\rm orb}$, and inclination angle constraints \citep{addison2021}.

In this study, we did not perform a full atmospheric retrieval mainly because only Fe~{\sc i} was ultimately detected thereby making the retrieved PT profile and metallicity parameters susceptible to strong degeneracies. However, based on the MCMC sampling analysis and the injection-recovery tests (Sect. \ref{sect:results}), we conclude that the observed Fe~{\sc i} emission spectrum has line strengths that are $\sim50\%$ lower than the nominal model shown in Fig. \ref{fig:template} (i.e., $\log_{10}\sim-0.3$, which is the value used for the injected signal that was chosen to roughly match the detected signal). This can roughly be achieved by, for example, reducing the temperature of the upper atmosphere from $4750\,{\rm K}$ to $4000\,{\rm K}$.

Comparable $\sim5\,{\rm km/s}$ blueshifts associated with various gaseous species have been detected by previous HRCC studies of UHJs \citep[e.g.,][]{snellen2010,line2021,gandhi2023}. For instance, detections of Fe~{\sc i} on the dayside of WASP-76~b using ESPRESSO show a blueshift of $-4.7\pm0.3\,{\rm km/s}$ \citep{costasilva2024}. Moreover, these authors report a higher blueshift during the pre-eclipse phase ($-6.0\pm0.4\,{\rm km/s}$) compared to the post-eclipse phases ($-3.3\pm0.5\,{\rm km/s}$). In the case of MASCARA-5~b, \citetalias{petz2025} do not report a $\Delta v$ value for their post-eclipse PEPSI observations to compare with our pre-eclipse EXPRES-based result; however, the Fe~{\sc i} detection shown in their Fig. 3 shows a $\Delta v\approx0\,{\rm km/s}$ suggesting that there could be a decrease in the blueshift between the pre- and post-eclipse phases.

The detected blueshift is most likely attributable to atmospheric dynamics such as winds, the planet's rotation, or magnetic drag, which is predicted to cause velocity shifts in $\Delta v$ and/or $K_{\rm p}$ \citep[e.g.,][]{zhang2017a,malsky2021,beltz2024,wardenier2025}. Neglecting 3D effects when generating the model templates used for cross-correlation can also lead to offsets on the order of a few ${\rm km/s}$ being inferred \citep{flowers2019a}, although these effects are more significant near quadrature when significant regions of both the nightside and dayside are visible \citep{beltz2020,beltz2022}. Ultimately, additional pre- and post-eclipse observations with a higher S/N will be needed to assess the origin of MASCARA-5~b's blueshifted Fe~{\sc i} emission lines and also to detect additional gaseous species on the planet's dayside. In particular, higher S/N measurements at optical wavelengths may reveal the presence of Fe~{\sc ii}, which will serve to better constrain the atmosphere's metallicity and thermal structure and provide clues about the nature of the inferred temperature inversion. At MASCARA-5~b's high atmospheric temperatures, CO likely exists at high abundance and may be detectable in the near IR in addition to H$_2$O and OH \citep[e.g.,][]{brogi2023,yan2023,smith2024a} using instruments like IGRINS-2 installed on Gemini North.

While the EXPRES instrument has been previously used to detect a planet's atmosphere through transmission spectroscopy \citep[Fe~{\sc i} was detected in MASCARA-2~b by][]{hoeijmakers2020}, this study is the first instance in which it has been used to successfully measure a planet's dayside emission spectrum. Therefore, this work serves as motivation for conducting similar studies of suitable UHJs in the future using instruments like EXPRES that are installed on relatively modest-sized telescopes \citep[e.g.,][]{pino2022}.

\acknowledgments{}
These results made use of the Lowell Discovery Telescope (LDT) at Lowell Observatory.  Lowell is a private, non-profit institution dedicated to astrophysical research and public appreciation of astronomy and operates the LDT in partnership with Boston University, the University of Maryland, the University of Toledo, Northern Arizona University and Yale University.

This work used the EXtreme PREcision Spectrograph (EXPRES) that was designed and commissioned at Yale with financial support by the U.S. National Science Foundation under MRI-1429365 and ATI-1509436 (PI D. Fischer).

J.S. acknowledges support from the GeoFamily Foundation and contributors to the Early Career Scientist Fund at Lowell Observatory.

J.S. and J.L. acknowledge support from National Science Foundation AAG award AST2009343.

E.B. was supported by the NSF REU program in astronomy and planetary science at Northern Arizona University (NSF award \#2349774).

This work is part of the research program VIDI New Frontiers in Exoplanetary Climatology with project number 614.001.601, which is (partly) financed by the Dutch Research Council (NWO).

We thank Peter Smith and Jake Kamen for helpful discussions.

\facilities{}
LDT(LMI, EXPRES)

\software{\texttt{AstroPy} \citep{astropycollaboration2013,astropycollaboration2018,astropycollaboration2022}, \texttt{matplotlib} \citep{Hunter:2007}, \texttt{numpy} \citep{harris2020array}, \texttt{scipy} \citep{2020SciPy-NMeth}, \texttt{petitRADTRANS} \citep{molliere2019,blain2024a}, \texttt{emcee} \citep{foreman-mackey2013,foreman-mackey2019}, \texttt{FastChem} \citep{stock2018}.}

\section{Appendix}\label{sect:appendix}

\begin{figure*}
	\centering
   \includegraphics[width=0.95\columnwidth]{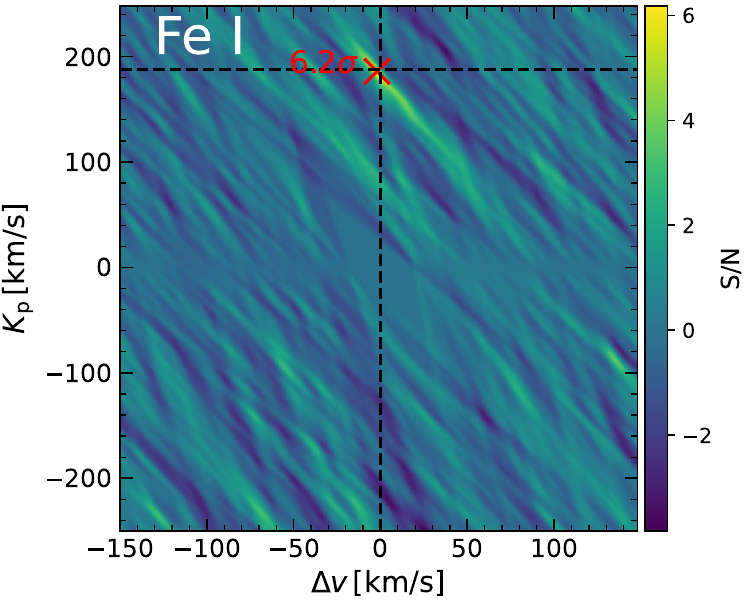}
   \includegraphics[width=0.95\columnwidth]{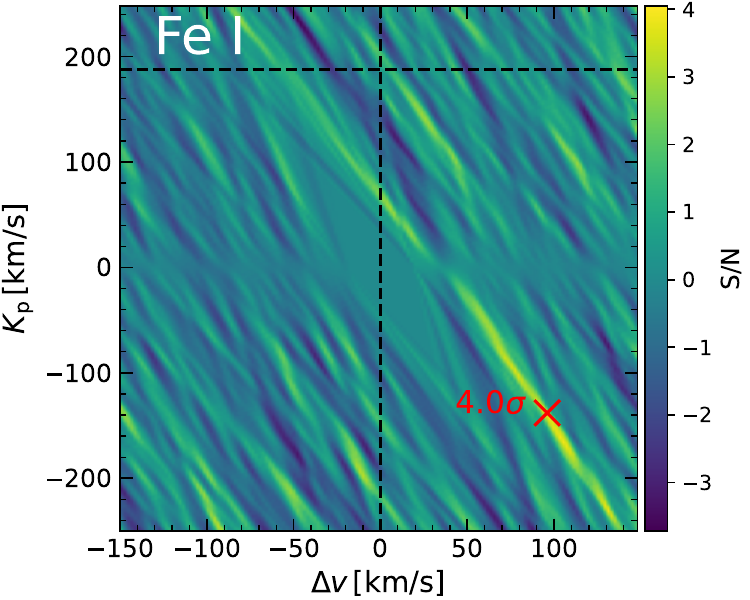}
   \includegraphics[width=0.95\columnwidth]{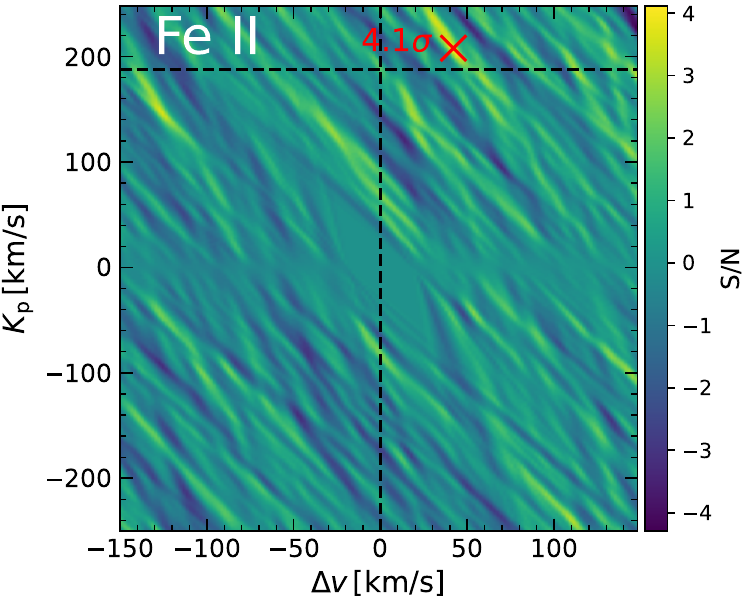}
   \includegraphics[width=0.95\columnwidth]{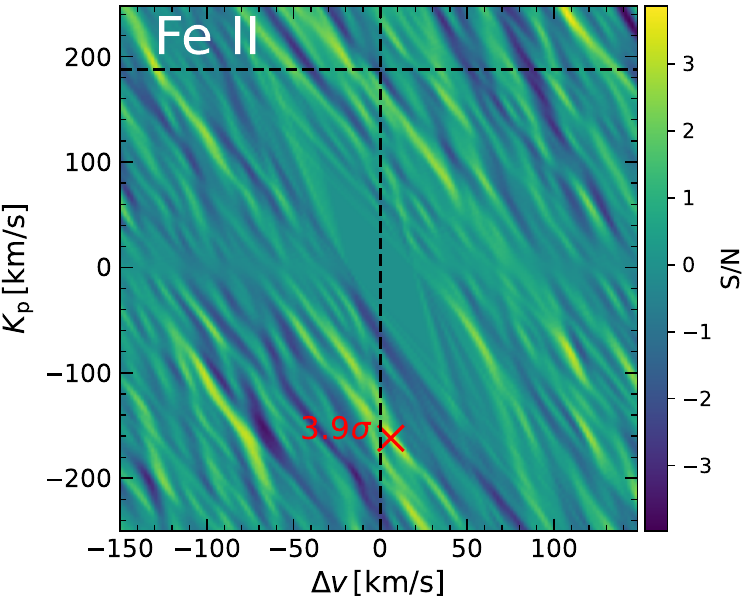}
   \includegraphics[width=0.95\columnwidth]{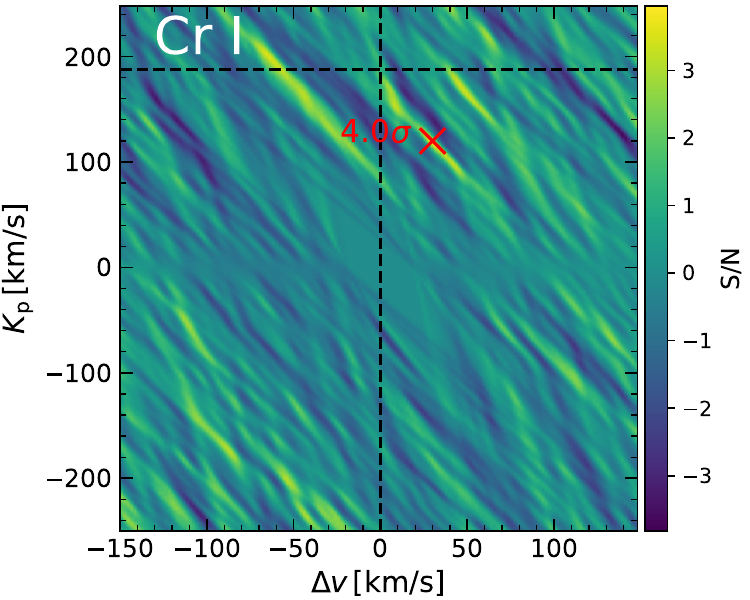}
   \includegraphics[width=0.95\columnwidth]{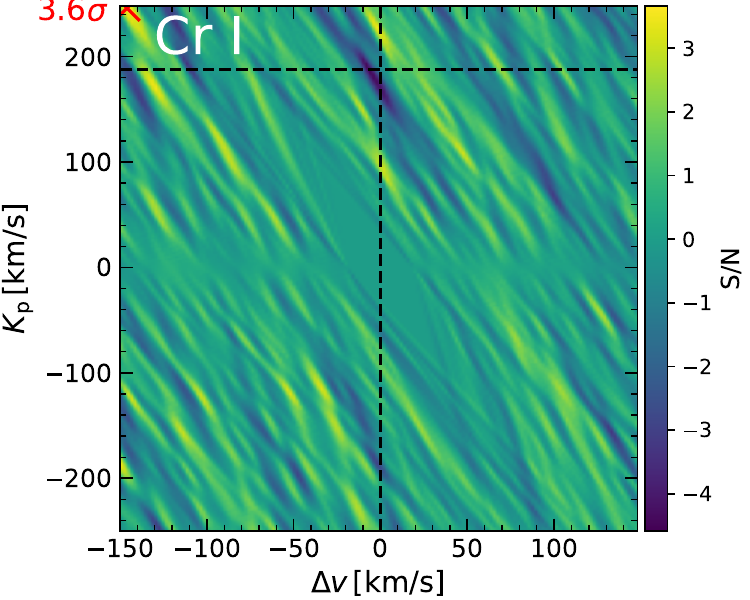}
	\caption{Same as Fig. \ref{fig:KpVsys} but calculated after masking the ${\rm CCF}(v,t)$ values with $|v|<20\,{\rm km/s}$ (i.e., near the strongest stellar residuals visible in Fig. \ref{fig:CCF} near $v=0\,{\rm km/s}$).}
	\label{fig:KpVsys_vlim20}
\end{figure*}

In Fig. \ref{fig:KpVsys_vlim20} we show the ${\rm CCF}(v,t)$ values calculated after first masking regions where $|v|<20\,{\rm km/s}$ in order to test how the final maps are impacted by residual stellar lines appearing in the cleaned spectra.

\bibliography{Hi-Res}{}
\bibliographystyle{aasjournal}

\end{document}

%% file: tables/published_param_table.tex
\begin{table}
       \caption{Adopted systemic parameters reported by \citep[$P_{\rm orb}$ and $T_0$][]{kokori2026} and \citep[all other parameters][]{addison2021} along with the values derived in this study.}\vspace{-0.3cm}
       \label{tbl:param}
       \begin{center}
       \begin{tabular}{@{\extracolsep{\fill}}l @{\hskip 1.cm} r @{\extracolsep{\fill}}}
              \hline
              \hline
              \noalign{\vskip0.5mm}
              \multicolumn{2}{c}{Published parameters} \\
              \hline
$T_{\rm eff}\,[{\rm K}]$       & $7690_{-250}^{+400}$   \\
$R_\star\,[R_\odot]$           & $1.92\pm0.07$          \\
$M_\star\,[M_\odot]$           & $1.90_{-0.08}^{+0.10}$ \\
$\log g_\star\,[{\rm cm/s^2}]$ & $4.15\pm0.04$          \\
$[{\rm M/H}]_\star$            & $0.090\pm0.030$        \\
$v\sin{i}_\star\,[{\rm km/s}]$ & $6.0\pm0.2$            \\
$R_{\rm p}\,[R_{\rm Jup}]$     & $1.49\pm0.05$ \\
$R_{\rm p}/R_\star$            & $0.07955_{-0.00053}^{+0.00063}$ \\
$M_{\rm p}\,[M_{\rm Jup}]$     & $3.12\pm0.18$ \\
$P_{\rm orb}\,[{\rm d}]$       & $2.65023153\pm0.000000380$ \\
$T_0\,[{\rm BJD}]^a$             & $2459558.098917\pm0.0000680$ \\
$T_{14}\,[{\rm hr}]$           & $2.489\pm0.009$ \\
$K_{\rm p}\,[{\rm km/s}]^a$    & $187.6\pm3.3$ \\
\noalign{\vskip0.5mm}
\hline
\multicolumn{2}{c}{Derived HRCC parameters} \\
\noalign{\vskip0.5mm}
\hline
$\log_{10}a^b$               & $-0.39_{-0.09}^{+0.08}$       \\
$K_{\rm p}\,[{\rm km/s}]$  & $187.1\pm3.4$ \\
$\Delta v\,[{\rm km/s}]$   & $-3.2\pm1.4$  \\
\noalign{\vskip0.5mm}
\hline
\multicolumn{2}{l}{$^a$Calculated from published parameters. $^\dagger$Note that $b$} \\
\multicolumn{2}{l}{scales the line strengths of the \citetalias{petz2025}-based model (Fig. \ref{fig:PT}).} \\
       \end{tabular}
       \end{center}
\end{table}


%% file: tables/inject_table.tex
\begin{table}
       \caption{Results of the injection/recovery test atmospheric retrieval. Both the injected model and the retrieval assumed a solar metallicity.}\vspace{-0.3cm}
       \label{tbl:inject}
       \begin{center}
       \begin{tabular}{@{\extracolsep{\fill}}l @{\hskip 1.cm} c r @{\extracolsep{\fill}}}
              \hline
              \hline
              \noalign{\vskip0.5mm}
Parameter & Injected & Recovered \\
\hline
$\log_{10}a$                     & $-0.3$   & $-0.43_{-0.11}^{+0.08}$   \\
$K_{\rm p}\,[{\rm km/s}]$        & $-190.4$ & $-192.9\pm3.2$   \\
$\Delta v\,[{\rm km/s}]$         & $0$      & $1.2\pm1.4$    \\
\noalign{\vskip0.5mm}
\hline
       \end{tabular}
       \end{center}
\end{table}